\documentclass[10pt,conference]{IEEEtran}
\IEEEoverridecommandlockouts
\usepackage{cite}
\usepackage{amsmath}
\usepackage{amsfonts}
\usepackage{bm}  
\usepackage{graphicx}
\usepackage[bookmarks=false]{hyperref}
\usepackage{amsthm} 
\usepackage{authblk} 
\usepackage{graphicx, amsmath, amsthm, amssymb, subcaption, url, cite, array, amsthm,booktabs,xcolor}

\def\BibTeX{{\rm B\kern-.05em{\sc i\kern-.025em b}\kern-.08em
    T\kern-.1667em\lower.7ex\hbox{E}\kern-.125emX}}
\makeatletter
\def\endthebibliography{%
	\def\@noitemerr{\@latex@warning{Empty `thebibliography' environment}}%
	\endlist
}

\begin{document}
\title{\huge Energy-Efficient Learning-Based Beamforming for ISAC-Enabled V2X Networks
}
\author[1]{Chen Shang}
\author{Jiadong Yu}
\author[2]{Dinh Thai Hoang}
\affil[1]{Thrust of Internet of Things, The Hong Kong University of Science and Technology (Guangzhou), China}
\affil[2]{School of Electrical and Data Engineering, University of Technology Sydney, Australia}
\vspace{-20pt}
\maketitle

\begin{abstract}
This work proposes an energy-efficient, learning-based beamforming scheme for integrated sensing and communication (ISAC)-enabled V2X networks. Specifically, we first model the dynamic and uncertain nature of V2X environments as a Markov Decision Process. This formulation allows the roadside unit to generate beamforming decisions based solely on current sensing information, thereby eliminating the need for frequent pilot transmissions and extensive channel state information acquisition.
We then develop a deep reinforcement learning (DRL) algorithm to jointly optimize beamforming and power allocation, ensuring both communication throughput and sensing accuracy in highly dynamic scenario.
To address the high energy demands of conventional learning-based schemes, we embed spiking neural networks (SNNs) into the DRL framework. Leveraging their event-driven and sparsely activated architecture, SNNs significantly enhance energy efficiency while maintaining robust performance.
Simulation results confirm that the proposed method achieves substantial energy savings and superior communication performance, demonstrating its potential to support green and sustainable connectivity in future V2X systems.

\end{abstract}

\vspace{-10pt}
\section{Introduction}
\vspace{-4pt}
The rapid evolution of wireless communication technologies has introduced promising paradigms such as integrated sensing and communication (ISAC) and artificial intelligence (AI), which together enable high-resolution environmental perception and intelligent decision-making in vehicle-to-everything (V2X) networks~\cite{10663823}. Within ISAC-assisted V2X systems, a key objective is to develop efficient beamforming strategies that enhance sensing accuracy, reduce beam training overhead, and thereby improve communication performance, i.e., enabling sensing-assisted communication~\cite{10944644}.

Recent ISAC-assisted beamforming methods can be broadly classified into two categories: optimization-based and learning-based approaches. Optimization-based techniques, such as Kalman filtering~\cite{liu2020radar,10561505}, rely on dynamic state estimation to adjust beam directions. However, these methods typically require accurate system models and known noise statistics, which limits their robustness and adaptability in dynamic and uncertain V2X scenarios. In contrast, learning-based approaches~\cite{liu2022learning,10605608,10304580} leverage deep learning or reinforcement learning to bypass explicit modeling, thereby significantly improving system performance under such conditions.

Despite their performance advantages, learning-based schemes remain highly energy-intensive and face significant deployment challenges in real-world V2X systems. These methods often rely on large neural networks with dense activations, leading to high computational demands during training, inference, and online adaptation~\cite{de2023growing}. For example, training a large-scale model such as GPT-3, which contains 175 billion parameters, consumed approximately 1,287 megawatt-hours (MWh) of electricity, comparable to the annual energy usage of 130 U.S. households~\cite{de2023growing}. Consequently, conventional learning-based strategies are unsuitable for energy-constrained roadside units (RSUs) that must operate continuously in dynamic, latency-sensitive environments. Furthermore, sustaining inference performance under fluctuating wireless conditions imposes ongoing energy burdens, conflicting with the low-power and sustainability requirements of next-generation green communication infrastructures.

To address these challenges, this work proposes a novel energy-efficient learning-based beamforming scheme for ISAC-enabled V2X system. To this end, we first model the dynamic and uncertain of vehicular network as a Markov Decision Process (MDP) framework. This framework eliminates the need for explicit CSI and significantly reduces beam training overhead, thereby improving system adaptability under dynamic channel and mobility conditions. To achieve model-free and adaptive control, we adopt a deep reinforcement learning (DRL) framework based on an actor-critic architecture with policy clipping. This design allows the RSU to learn effective policies directly through interaction with the environment, without relying on model linearization or prior knowledge of system noise statistics. Most importantly, to enhance energy efficiency, we integrate Spiking Neural Networks (SNNs) into the DRL framework. Leveraging their sparse and event-driven computation, SNNs significantly reduce energy consumption while maintaining fast inference and high-quality decision-making, making them well-suited for resource-constrained and latency-sensitive V2X scenarios. \textit{To the best of our knowledge, this is the first work to leverage SNN-driven DRL for green ISAC communication systems.} Extensive simulations validate the superiority of the proposed energy-efficient learning-based algorithm in terms of both communication performance and energy consumption.

\vspace{-5pt}
\section{System Model and Problem Formulation}\label{Se:system model}
\begin{figure}[t!]
    \centering
    \includegraphics[width=0.48\textwidth]{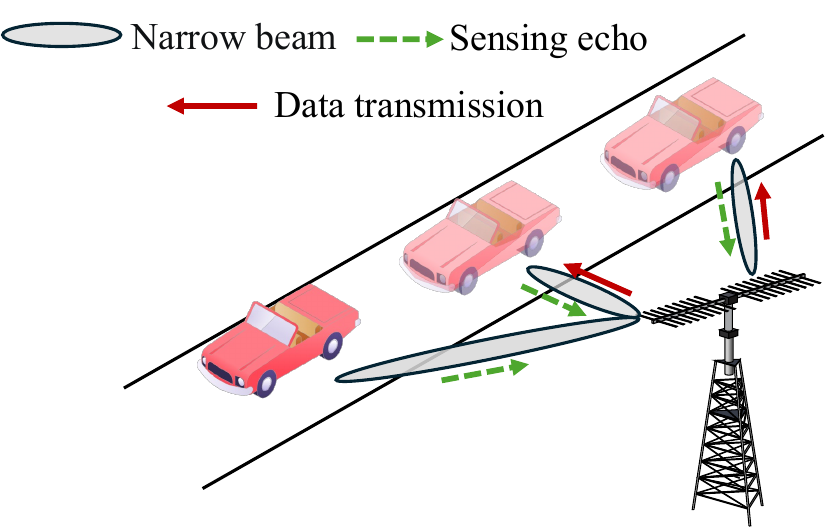}
    \caption{An AI-powered ISAC-assisted V2X system. The RSU leverages SNN-based decision-making to reduce energy consumption.}
    \label{Fig: scenario}
    \vspace{-5pt}
\end{figure}

\vspace{-5pt}
As shown in Fig.~\ref{Fig: scenario}, we consider an ISAC-assisted V2X network comprising an RSU and multiple vehicles $\mathcal{K} = \{1, 2, \ldots, K\}$ traveling along a straight single-lane road parallel to the RSU's antenna array. The RSU equipped with massive MIMO uniform linear arrays (ULA) including $N_{\text{TA}}$ transmit and $N_{\text{RA}}$ receive antennas, transmits ISAC signals that integrate both radar (sensing) and communication components to vehicles equipped with single antennas.
\vspace{-8pt}
\subsection{Sensing and Communication Model}\label{sensing model}
\vspace{-5pt}
Let $x_{k,n}(t)$ denote the ISAC signal for vehicle-$k$ at time $t$ in the $n$-th time slot ($n \in \{1, 2, \ldots, N\}$). The transmitted signal vector is $\boldsymbol{x}_n(t) = [x_{1,n}(t), \ldots, x_{K,n}(t)]^T \in \mathbb{C}^{K \times 1}$. The signal sent through the RSU's transmit array is given by:
\begin{equation}
\tilde{\boldsymbol{x}}_n(t) = \mathbf{F}_n \boldsymbol{x}_n(t) \in \mathbb{C}^{N_{\text{TA}} \times 1},
\end{equation}
where $\mathbf{F}_n = [\mathbf{f}_{1,n}, \ldots, \mathbf{f}_{K,n}]$ is the transmit beamforming matrix, and $\mathbf{f}_{k,n} \in \mathbb{C}^{N_{\text{TA}} \times 1}$ denotes the beamformer for vehicle-$k$. Accordingly, the received echo at the RSU can be expressed as:
\begin{equation}
\begin{aligned}
\boldsymbol{y}_n(t) = ~&\mathcal{E} \sum_{k=1}^K \sqrt{p_{k,n}} \beta_{k,n} e^{j2\pi \mu_{k,n} t} \mathbf{b}(\theta_{k,n}) \mathbf{a}^H(\theta_{k,n}) \\
&\tilde{\boldsymbol{x}}_n(t - \tau_{k,n}) + \mathbf{z}(t),
\end{aligned}
\label{sumisac}
\end{equation}
where $\mathcal{E} = \sqrt{N_{\text{TA}} N_{\text{RA}}}$ is the array gain and $p_{k,n}$ is the RSU’s transmit power for beam-$k$. $\beta_{k,n} = \kappa / (2 d_{k,n})$ is the reflection coefficient with $\kappa$ as the fading factor and $d_{k,n}$ the distance between vehicle-$k$ and the RSU. Moreover, $\mu_{k,n}$, $\tau_{k,n}$, and $\theta_{k,n}$ represent the Doppler shift, round-trip delay, and angle of arrival (AoA), respectively. $\mathbf{z}(t) \sim \mathcal{CN}(0, \sigma^2 \mathbf{I})$ denotes additive white Gaussian noise (AWGN). Moreover, ${\mathbf{a}}\left( {{\theta_{k,n}}} \right)$ and ${\mathbf{b}}\left( {{\theta_{k,n}}} \right)$ are the transmit and receive steering vectors of the RSU antenna array, which are given by:
\begin{equation}
\mathbf{a}(\theta_{k,n}) = \sqrt{\frac{1}{N_{\text{TA}}}} \left[1, e^{-j\pi \cos \theta_{k,n}}, \ldots, e^{-j\pi (N_{\text{TA}} - 1) \cos \theta_{k,n}} \right]^T,
\label{eq4}
\end{equation}
\begin{equation}
\mathbf{b}(\theta_{k,n}) = \sqrt{\frac{1}{N_{\text{RA}}}} \left[1, e^{-j\pi \cos \theta_{k,n}}, \ldots, e^{-j\pi (N_{\text{RA}} - 1) \cos \theta_{k,n}} \right]^T.
\label{eq5}
\end{equation}
Note that the received echo signal consists of multiple sub-echoes originating from different vehicles. To isolate the individual signal corresponding to each vehicle, the RSU can apply massive MIMO theory~\cite{marzetta2016fundamentals} and spatial filtering techniques~\cite{richards2005fundamentals}, as expressed by:
\begin{equation}
\begin{aligned}
y_{k,n}\left( t \right) =&\mathbf{b}^H( \hat{\theta}_{k,n} ) \boldsymbol{y}_{n}\left( t \right) 
\\
=&\mathcal{E} \sqrt{p_{k,n}}\beta _{k,n}e^{j2\pi \mu _{k,n}t}\mathbf{a}^H\left( \theta _{k,n} \right) \tilde{\boldsymbol{x}}_n\left( t-\tau _{k,n} \right) 
\\
&+z_{k,n}\left( t \right) 
\end{aligned},
\label{eq:echo for vehicle}
\end{equation}
where $\mathbf{b}^H( \hat{\theta}_{k,n}) \boldsymbol{y}_{n}\left( t \right) $ represents the spatial filtering operation based on the estimated angle $\hat{\theta}_{k,n}$ ($\hat{\theta}_{k,n}\approx{\theta}_{k,n}$) of the receive beamforming vector. $z_{k,n}(t) = \mathbf{b}^H( \hat{\theta}_{k,n} ) \mathbf{z}(t)$ with $z_{k,n}\left( t \right) \sim \mathcal{C} \mathcal{N} \left( 0,\sigma _{z}^{2} \right) $ denotes the noise vector~\cite{11091493}.

After extracting the echo signals corresponding to each vehicle, the RSU estimates their motion states and updates the beamforming strategy accordingly. To improve signal detectability, matched filtering is employed to maximize the signal-to-noise ratio (SNR), enabling the extraction of the Doppler frequency $\tilde{\mu}_{k,n}$ and time delay $\tilde{\tau}_{k,n}$ from the received signal~\cite{liu2022learning}. Then, the measured time delay and Doppler frequency relate to the estimated distance and velocity as:
\begin{equation}
\tilde{\tau}_{k,n} = \frac{2\hat{d}_{k,n}}{c} + z_{\tau_k,n},\quad 
\tilde{\mu}_{k,n} = \frac{2\hat{v}_{k,n} \cos \theta_{k,n} f_c}{c} + z_{\mu_k,n},
\label{measure}
\end{equation}
where $c$ and $f_c$ are the speed of light and carrier frequency, respectively. The noise terms $z_{\tau_k,n}$ and $z_{\mu_k,n}$ are zero-mean with variances $\sigma^2_{\tau_k,n}$ and $\sigma^2_{\mu_k,n}$, respectively, which are inversely proportional to the received signal-to-interference-plus-noise ratio (SINR)~\cite{liu2020radar}:
\begin{equation}
\mathrm{SINR}_{k,n} = \frac{\mathcal{E}^2 p_{k,n} |\beta_{k,n}|^2 |\mathbf{a}^H(\theta_{k,n}) \mathbf{f}_{k,n}|^2}
{\sum_{i \ne k} \mathcal{E}^2 p_{i,n} |\beta_{i,n}|^2 |\mathbf{a}^H(\theta_{k,n}) \mathbf{f}_{i,n}|^2 + \sigma_z^2}.
\label{eq:sensing sinr}
\end{equation}
Accordingly, the noise variances are expressed as:
\begin{equation}
\sigma^2_{\tau_k,n} = \frac{\alpha^2_{\tau} \left( \sum_{i \ne k} \mathcal{E}^2 p_{i,n} |\beta_{i,n}|^2 |\mathbf{a}^H(\theta_{k,n}) \mathbf{f}_{i,n}|^2 + \sigma_z^2 \right)}
{\mathcal{E}^2 p_{k,n} |\beta_{k,n}|^2 |\mathbf{a}^H(\theta_{k,n}) \mathbf{f}_{k,n}|^2},
\end{equation}
\begin{equation}
\sigma^2_{\mu_k,n} = \frac{\alpha^2_{\mu} \left( \sum_{i \ne k} \mathcal{E}^2 p_{i,n} |\beta_{i,n}|^2 |\mathbf{a}^H(\theta_{k,n}) \mathbf{f}_{i,n}|^2 + \sigma_z^2 \right)}
{\mathcal{E}^2 p_{k,n} |\beta_{k,n}|^2 |\mathbf{a}^H(\theta_{k,n}) \mathbf{f}_{k,n}|^2},
\end{equation}
where $\alpha^2_{\tau}$ and $\alpha^2_{\mu}$ are constants determined by system configuration and signal design.

It is worth noting that the RSU operates as a mono-static radar when no vehicles are within its coverage area. Once a vehicle enters the coverage zone, the RSU estimates its state and promptly performs beam alignment for data transmission. Given the beamforming vector $\mathbf{f}_{k,n}$ and transmit power $p_{k,n}$, the received communication signal at vehicle-$k$ in time slot $n$, denoted by $\mathcal{C}_{k,n}(t)$, is expressed as:
\begin{equation}
\begin{aligned}
\mathcal{C}_{k,n}(t) = &~\bar{\mathcal{E}}\sqrt{p_{k,n}} \alpha_{k,n} e^{j2\pi \mu'_{k,n} t} \mathbf{a}^H(\theta_{k,n}) 
\sum_{k=1}^K \mathbf{f}_{k,n} x_{k,n}(t) \\
&+ z_c(t),
\end{aligned}
\label{eq:communication for vehicle}
\vspace{-8pt}
\end{equation}
where $\bar{\mathcal{E}} = \sqrt{N_{\text{TA}}}$ is the communication array gain, and $\alpha_{k,n} = \sqrt{\alpha_0 (d_{k,n}/d_0)^{-\varrho}}$ denotes the large-scale fading coefficient, with $\alpha_0$, $d_0$, and $\varrho$ representing the reference path loss, distance, and path loss exponent, respectively. The term $z_c(t) \sim \mathcal{CN}(0, \sigma_c^2)$ denotes complex Gaussian noise.
Accordingly, the communication SINR at vehicle-$k$ is given by:
\begin{equation}
\gamma_{k,n}(\mathbf{f}_{k,n}, p_{k,n}) = 
\frac{\bar{\mathcal{E}}^2 p_{k,n} |\alpha_{k,n} \mathbf{a}^H(\theta_{k,n}) \mathbf{f}_{k,n}|^2}
{\sum_{i \ne k} \bar{\mathcal{E}}^2 p_{i,n} |\alpha_{k,n} \mathbf{a}^H(\theta_{k,n}) \mathbf{f}_{i,n}|^2 + \sigma_c^2}.
\label{SINR of communication}
\vspace{-2pt}
\end{equation}
Therefore, the achievable sum-rate of the ISAC-assisted V2X system is then expressed as:
\begin{equation}
R = \sum_{k=1}^K R_k = \sum_{k=1}^K \log_2\left(1 + \gamma_{k,n}(\mathbf{f}_{k,n}, p_{k,n})\right).
\label{sum-rate}
\end{equation}

As observed in ~\eqref{sum-rate}, the achievable rate depends heavily on the beamforming and power allocation schemes. Notably, accurate sensing improves beam alignment quality, enabling more effective sensing-assisted communication. However, since the true vehicle states are unknown, the RSU cannot directly evaluate its sensing accuracy via metrics like Root-mean-square deviation (RMSE). To overcome this limitation, we introduce the Cramér-Rao Lower Bound (CRLB) as a performance benchmark~\cite{kay1993fundamentals}, which 
provides a theoretical lower bound on the variance of any unbiased estimator in target parameter estimation. 

According to CRLB theory~\cite{kay1993fundamentals}, given the estimated angle $\hat{\theta}_{k,n}$ and distance $\hat{d}_{k,n}$, their estimation error is bounded as:
\begin{equation}
\mathbb{E} \left[ \left( \hat{\theta}_{k,n} - \theta_{k,n} \right)^2 \right] \geq \mathrm{CRLB}_{\theta}(\theta_{k,n}, \mathbf{f}_{k,n}, p_{k,n}),
\end{equation}
\begin{equation}
\mathbb{E} \left[ \left( \hat{d}_{k,n} - d_{k,n} \right)^2 \right] \geq \mathrm{CRLB}_{d}(d_{k,n}, \mathbf{f}_{k,n}, p_{k,n}).
\vspace{-6pt}
\end{equation}
The closed-form expressions of $\mathrm{CRLB}_{\theta}$ and $\mathrm{CRLB}_{d}$ are given by~\cite{9652071}:
\begin{equation}
\mathrm{CRLB}_{\theta}(\theta_{k,n}, \mathbf{f}_{k,n}, p_{k,n}) = \left[ \frac{1}{\sigma_{y_k}^2} \left( \frac{\partial \tilde{y}_{k,n}}{\partial \theta_{k,n}} \right)^H \frac{\partial \tilde{y}_{k,n}}{\partial \theta_{k,n}} \right]^{-1},
\label{eq:crlb-theta}
\end{equation}
\begin{equation}
\mathrm{CRLB}_d(d_{k,n}, \mathbf{f}_{k,n}, p_{k,n}) = \left[ \frac{1}{\sigma^2_{\tau_k}} \left( \frac{2}{c} \right)^2 \right]^{-1},
\vspace{-5pt}
\end{equation}
where $\tilde{y}_{k,n} = \mathcal{E} \sqrt{p_{k,n}} \beta_{k,n} \xi \mathbf{a}^H(\theta_{k,n}) \mathbf{f}_{k,n} + \tilde{z}_{k,n}$ is the matched-filtered signal with gain $\xi$.

\vspace{-5pt}
\subsection{Problem Formulation}
\vspace{-5pt}
As illustrated above, the RSU aims to maximize the sum-rate while ensuring sensing accuracy through joint optimization of beamforming and power allocation. Thus, the optimization problem is formulated as follows:
\begin{subequations}\label{eqn:optimization-problem}
\begin{align}
  \textbf{P1:}~ &\underset{\{\mathbf{F}_n,~\boldsymbol{p}_n\}_{n=1}^{N}}{\max}\,\mathbb{E} \left[ \frac{1}{N}\sum_{n=1}^N{\sum_{k=1}^K{\log _2\left( 1+\gamma _{k,n}\left( \mathbf{f}_{k,n},p_{k,n} \right) \right)}} \right] , \tag{\ref{eqn:optimization-problem}} \\
\text{s.t.}\quad & \mathbb{E} \left[ \frac{1}{N}\frac{1}{K}\sum_{k=1}^K{\mathrm{CRLB}_{\theta}\left( \theta _{k,n},\mathbf{f}_{k,n},p_{k,n} \right)} \right] \leqslant \epsilon _{\theta}, \label{eqn:pbm-const-1} \\
& \mathbb{E} \left[ \frac{1}{N}\frac{1}{K}\sum_{k=1}^K{\mathrm{CRLB}_{d}\left( d_{k,n},\mathbf{f}_{k,n},p_{k,n} \right)} \right] \leqslant \epsilon _d, \label{eqn:pbm-const-2} \\
& \sum_{k=1}^K{p_{k,n}}\leqslant P_{\max} ~~\text{and} ~~p_{k,n}>0 ~~~\forall k, \label{eqn:pbm-const-3}
\end{align} 
\end{subequations}
where $\boldsymbol{p}_n=\left[ p_{1,n},p_{2,n},\dots ,p_{K,n} \right] ^T\in \mathbb{C} ^{K\times 1}$ is the allocated power vector, $\epsilon _\theta$ and $\epsilon _d$ are the maximum tolerable CRLB thresholds for reliable sensing, and $P_{\max}$ represents the RSU's maximum transmit power.

Optimizing \textbf{P1} poses significant challenges for conventional methods. For instance, Kalman-based approaches such as the EKF require accurate system models and known noise statistics~\cite{liu2020radar,10561505}, which are difficult to obtain in dynamic and uncertain V2X environments. Moreover, their reliance on linearization introduces estimation errors under nonlinear vehicle dynamics and fast-varying channel conditions. Although learning-based methods such as ~\cite{liu2022learning,10605608,10304580} have shown superior adaptability and performance, they often suffer from high energy consumption due to the computational complexity of deep neural networks during training and inference. This makes them impractical for deployment in power-constrained RSUs that operate continuously under real-time requirements.

\vspace{-7pt}
\section{Proposed Learning Algorithm}\label{sec3}
\vspace{-7pt}
\subsection{Model-free DRL Algorithm}\label{mdp}
To address above challenges, we propose an energy-efficient, model-free DRL framework based on Spiking Neural Networks (SNNs). This framework enables the RSU to learn optimal policies from sensing data while significantly reducing energy consumption, thereby achieving adaptive and sustainable ISAC performance in V2X networks. To this end, we first model the dynamic and uncertain environment of V2X as a Markov Decision Process (MDP), defined by the tuple $\left( \mathcal{S}, \mathcal{A}, \mathcal{P}, r, G \right)$, where $\mathcal{S}$ and $\mathcal{A}$ denote the state and action spaces, $\mathcal{P}$ represents the state transition probability, $r$ is the reward function, and $G \in (0,1)$ is the discount factor for long-term return. The MDP elements are defined as follows:

\subsubsection{State Space}
At each time slot $n$, the system state is defined as:
\begin{equation}
\mathcal{S} = \left\{ \left\{ \hat{\theta}_{k,n}, \hat{d}_{k,n}, \hat{v}_{k,n}, \hat{\gamma}_{k,n} \right\}; ~ 1 \leq k \leq K \right\}.
\end{equation}
Note that all state variables are subject to estimation errors due to sensing imperfections. These errors directly impact the learning performance. Nevertheless, the proposed DRL algorithm does not require perfect state information and remains robust to estimation inaccuracies, enabling near-optimal performance even under imperfect sensing.

\subsubsection{Action Space and Reward Function}
At each time slot $n$, the RSU takes an action from the action space $\mathcal{A}$, i.e., it determines the beamforming vectors and power allocation for all vehicles based on the observed state. The action space is defined as:
\begin{equation}
\mathcal{A} = \left\{ \mathbf{F}_n, \boldsymbol{p}_n \right\}.
\vspace{-5pt}
\end{equation}

Upon executing an action $a_n \in \mathcal{A}$ in state $s_n$, the RSU receives an immediate reward $r_n(a_n, s_n)$, which reflects the system's performance in communication and sensing. Given the optimization goals in (\ref{eqn:optimization-problem}), the reward function is defined as:
\begin{equation}
r_n(a_n, s_n) = \mathbf{1}(R \cdot \mathcal{J} - \mathrm{CRLB}_\theta - \mathrm{CRLB}_d),
\label{eq:reward function}
\end{equation}
where $\mathbf{1}(\cdot)$ is an indicator function that equals 1 if constraints (\ref{eqn:pbm-const-1})–(\ref{eqn:pbm-const-3}) are satisfied, and 0 otherwise. Here, $R$ is the achievable sum-rate defined in (\ref{sum-rate}), and $\mathcal{J}$ denotes the Jain's fairness index, given by:
\begin{equation}
\mathcal{J} \left( R_1,R_2,\dots ,R_K \right) =\frac{2\left( \sum_{k=1}^K{R_k} \right) ^2}{K\times \sum_{k=1}^K{R_{k}^{2}}}.
\end{equation}

\begin{figure}[t]
    \centering
    \includegraphics[width=0.51\textwidth]{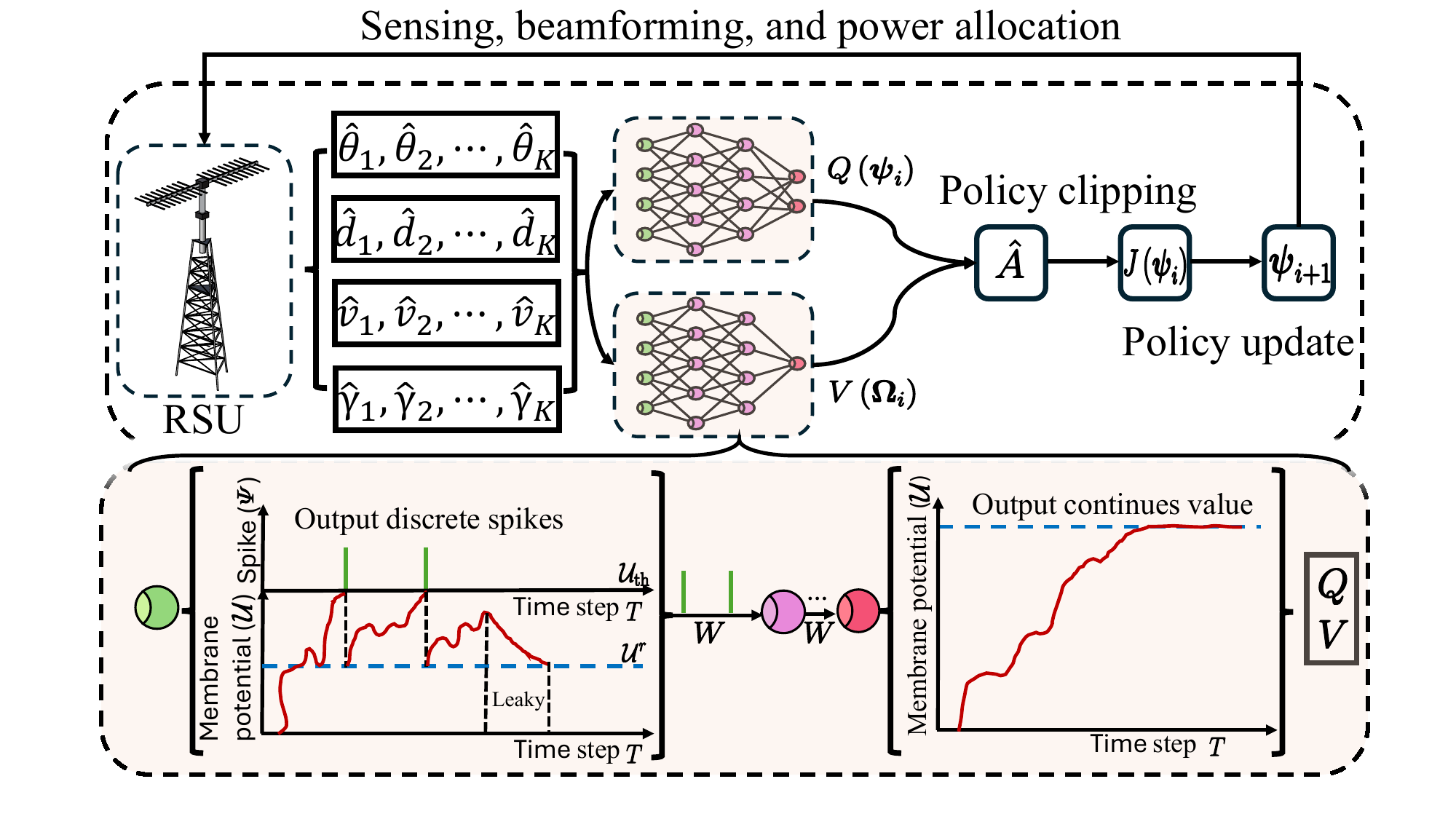}
    \caption{The illustration of the training process for the proposed SNNs-driven DRL.}
    \label{Fig:overall}
    \vspace{-5pt}
\end{figure}
This reward is designed to guide the RSU to jointly optimize beamforming and power allocation, enhancing both sum-rate and fairness. Additionally, it promotes exploration of sensing-aware strategies by embedding CRLB-based penalties, effectively balancing communication rate and sensing precision.

Leveraging the MDP framework, we are able to reformulate the objective function and develop the DRL algorithm. Let $\pi$ denote a stochastic policy of the RSU (i.e., $\pi :\mathcal{S} \times \mathcal{A} \rightarrow \left[ 0,1 \right] $), which is the probability that taking action $a_n$ given the state $s_n$, i.e., $\pi =\mathbb{P} \left\{ a_n\left| s_n \right. \right\} $. In particular, the objective of RSU is to maximize the cumulative reward (i.e., communication rate and sensing accuracy) over time. Therefore, given the discount of long-term reward $G$,  the expected discounted reward of the RSU follows policy $\pi $ is given by:
\begin{equation}
J\left( \pi \right) =\mathbb{E} _{a_n\sim \pi ,s_n\sim \mathcal{P}}\left[ \sum_{n=0}^{\infty}{G^nr_n\left( s_n,a_n \right)} \right],
\end{equation}
where $\mathcal{P} \left( s_{n+1}\left| s_n,a_n \right. \right) $ is the state transition probability distribution that models the dynamics of the environment and is unknown to the RSU. As a result, the optimization problem in \textbf{P1} can be transformed into finding the optimal policy \(\pi ^*\) that maximizes \(J(\pi )\), expressed as:
\begin{subequations}\label{eqn:maximization-problem}
\begin{align}
&\textbf{P2:}~~~~\mathrm{arg}\underset{\pi}{\max}\,J(\pi )\tag{\ref{eqn:maximization-problem}} \\ 
&\text{s.t.}~ a_n \sim \pi (a_n | s_n), s_{n+1} \sim \mathcal{P}(s_{n+1} | s_n, a_n).  \label{eqn:policy-and-transition}
\end{align}
\label{eq10}
\vspace{-7pt}
\end{subequations}

The proposed learning algorithm is illustrated in~Fig.~\ref{Fig:overall}. Specifically, as shown at the top~Fig.~\ref{Fig:overall}, the learning-based algorithm employs two specialized neural networks to optimize the RSU's operations: an actor network and a critic network. The actor network, parameterized by $\boldsymbol{\psi}$ (comprising weights and biases), generates the policy that guides the RSU’s beamforming and power allocation strategies. Concurrently, the critic network, denoted as $\boldsymbol{\Omega}$, evaluates the performance of the current policy by estimating the state-value function. Note that these two neural networks are driven by SNN rather than conventional neural networks, which are detailed in the Section~\ref{algorithm}. This evaluation serves as feedback to optimize $\boldsymbol{\psi}$, thereby enhancing the accuracy and efficiency of decision-making. Accordingly, the optimal policy in (\ref{eq10}) can be approximated as $\pi^* \gets \pi_{\boldsymbol{\psi}}$, where $\pi_{\boldsymbol{\psi}}(a_n \mid s_n) = \mathbb{P}(a_n \mid s_n; \boldsymbol{\psi})$. Since we focus on per-step behavior decisions, the time slot index $n$ is omitted hereafter for clarity.
Once the outputs of the two networks are obtained, an advantage function $\hat{A}(\cdot)$ is introduced to evaluate the current policy and guide its update. Specifically, the advantage function quantifies how much better the selected action is compared to the average policy behavior, and is defined as~\cite{schulman2015high}:
\begin{equation}
\hat{A}(s, a; \boldsymbol{\psi}) = Q(s, a; \boldsymbol{\psi}) - V(s; \boldsymbol{\Omega}),
\label{advanf}
\end{equation}
where $Q(s, a; \boldsymbol{\psi}) = \mathbb{E}_{a \sim \pi_{\boldsymbol{\psi}}, s \sim \mathcal{P}} \left[ \sum_{l=0}^{\infty} G^l r(s_l, a_l) \right]$ is the action-value function, $V(s; \boldsymbol{\Omega}) = \mathbb{E}_{s \sim \mathcal{P}} \left[ \sum_{l=0}^{\infty} G^l r(s_l, a_l) \right]$ is the state-value function. Here, and $l$ is the constant related to time step.
Based on $\hat{A}(s, a; \boldsymbol{\psi})$, the objective function of the actor network is defined as~\cite{schulman2017proximal}:
\begin{equation}
J(\boldsymbol{\psi}) = \min \left( \frac{\pi_{\boldsymbol{\psi}}(a \mid s)}{\pi_{\boldsymbol{\psi}_{\text{old}}}(a \mid s)} \hat{A},~ \phi(\epsilon, \hat{A}) \right),
\label{eq:loss of actor}
\end{equation}
where $\pi_{\boldsymbol{\psi}_{\text{old}}}$ denotes the policy parameters before update. The clipping function $\phi(\epsilon, \hat{A})$ is used to constrain the update ratio within a predefined range, thus ensuring stable gradient steps, which is given by:
\begin{equation}
\phi(\epsilon, \hat{A}) = 
\begin{cases}
(1 + \epsilon) \hat{A}, & \text{if } \hat{A} \geq 0, \\
(1 - \epsilon) \hat{A}, & \text{if } \hat{A} < 0.
\end{cases}
\label{eq:policy-clipping}
\end{equation}
This clipping mechanism stabilizes the training process by preventing overly aggressive policy changes and improves convergence performance.

As aforementioned, the complexity and frequent computations of conventional neural networks impose significant energy demands on the RSU. To address these challenges, we propose an enhanced DRL algorithm driven by SNNs in the following.

\subsection{Energy-efficient DRL-enabled by SNN}\label{algorithm}
Different conventional DRL algorithms, we replace neural networks with SNNs to reduce computational complexity and energy consumption. Unlike conventional networks that use continuous activations and multiply-and-accumulate (MAC) operations, SNNs operate in an event-driven manner by propagating discrete binary spikes. This allows training and inference using accumulate (AC) operations, while efficiently capturing temporal dependencies. The details of the SNN are illustrated at the bottom of Fig.~\ref{Fig:overall} and further elaborated below.
\subsubsection{SNN Framework}
We adopt the Leaky Integrate-and-Fire (LIF) model~\cite{yao2022glif} to construct the SNNs. Within SNNs, neuron-$j$ maintains a membrane potential $\mathcal{U}_j(\tilde{t})$ at SNN time step $\tilde{t} \ll n$, its dynamics can be expressed as:
\begin{equation}
\mathcal{U}_j(\tilde{t}+1) = (1-\lambda)\mathcal{U}_j(\tilde{t}) + \lambda \sum_{i=1}^M W_{i,j} \mathcal{I}_i(\tilde{t}),
\label{LIF}
\end{equation}
where $\lambda$ is the leakage coefficient, $W_{i,j}$ is the synaptic weight, and $\mathcal{I}_i(\tilde{t})$ is the input from presynaptic neuron $i$. As shown at the bottom of Fig.~\ref{Fig:overall}, all inputs are directly encoded into the membrane potential $\mathcal{U}_j(\tilde{t})$. When $\mathcal{U}_j(\tilde{t})$ exceeds the firing threshold $\mathcal{U}_{\text{th}}$, the neuron emits a spike $\varPsi_j = 1$ and resets its potential to $ \mathcal{U}_{\text{th}}$. Otherwise, $\varPsi_j = 0$ and the potential remains unchanged at time step $\tilde{t}+1$. In addition, we remove the firing threshold $\mathcal{U}_{\text{th}}$ in the final layer, allowing it to output continuous membrane potentials for both the action and value prediction, while maintaining spiking behavior in earlier layers.

\subsubsection{Training of SNN-Driven DRL}
Training neural networks requires calculating gradients through back-propagation, which relies on the differentiability of the loss function. 
Specifically, due to recurrent temporal dynamics in $\mathcal{U}_j(\tilde{t})$, gradient computation must unfold across time. Therefore, the back-propagation gradient with respect to synaptic weights is given by~\cite{11091493}:
\begin{equation}
\begin{aligned}
	&\frac{\partial J}{\partial W_{i,j}}=\sum_{\tilde{t}=1}^{\tilde{T}}{\frac{\partial J}{\partial \mathcal{U} _j\left( \tilde{t} \right)}\frac{\partial \mathcal{U} _j\left( \tilde{t} \right)}{\partial W_{i,j}}}\\
	&=\sum_{\tilde{t}=1}^{\tilde{T}-1}{\left( \frac{\partial J}{\partial \varPsi _j\left( \tilde{t} \right)}\frac{\partial \varPsi _j\left( \tilde{t} \right)}{\partial \mathcal{U} _j(\tilde{t})}+\frac{\partial J}{\partial \mathcal{U} _j(\tilde{t}+1)}\frac{\partial \mathcal{U} _j(\tilde{t}+1)}{\partial \mathcal{U} _j(\tilde{t})} \right)} \\
	&~\quad \times \frac{\partial \mathcal{U} _j(\tilde{t})}{\partial W_{i,j}}+\frac{\partial J}{\partial \varPsi _j( \tilde{T} )}\frac{\partial \varPsi _j( \tilde{T} )}{\partial \mathcal{U} _j(\tilde{t})}\frac{\partial \mathcal{U} _j(\tilde{t})}{\partial W_{i,j}}\\
\end{aligned},
\label{eq.bp}
\end{equation}
where $\tilde{T}$ is the total number of SNN time steps. It can be observed that the spiking function $\varPsi_j$ is non-differentiable due to its binary nature and Dirac delta derivative. To overcome this, we adopt a surrogate gradient method~\cite{10636728}, enabling effective learning in temporal environments. In particular, we replace $\varPsi_j$ with a smooth surrogate $\varphi(\varPsi)$:
\begin{equation}
\varphi(\varPsi) = \frac{1}{\pi} \arctan\left(\frac{\pi \eta}{2} \varPsi\right) + \frac{1}{2},
\label{ada}
\end{equation}
\begin{equation}
\varphi ^{\prime}\left( \varPsi \right) =\frac{\eta}{2}\times \frac{1}{1+\left( \frac{\pi \eta}{2}\varPsi \right) ^2}.
\label{ada1}
\end{equation}
where $\eta$ is a tunable smoothing parameter and $\varphi'(\varPsi)$ denotes the derivative of $\varphi(\varPsi)$. Note that the surrogate function approximate the spiking behavior during training (i.e., back-propagation), while the forward dynamics still follow the original LIF model in (\ref{LIF}). As a result, the surrogate-based approach enables effective training of SNN-driven DRL policies by providing meaningful gradients during back-propagation.

\vspace{-5pt}
\section{Performance Evaluations} \label{PE}
\vspace{-5pt}
\subsection{Simulation Settings}\label{simulation setting}
\textit{1) V2X Network:} We consider a V2X scenario with $K = 3$ single-antenna vehicles moving along a straight road and entering the coverage of an RSU located at (0, 0), equipped with $N_{\text{TA}} = N_{\text{RA}} = 32$ antennas and operating at $f_c = 30$~GHz~\cite{liu2020radar}. The initial vehicle positions are set to $(-5, 10)$, $(-15, 10)$, and $(-25, 10)$, respectively. Vehicle speeds follow $v \sim \mathrm{Unif}(10~\text{m/s}, 14~\text{m/s})$. The evolution noise is configured as $\sigma_{\theta} = 0.02^{\circ}$, $\sigma_d = 0.2$~m, and $\sigma_v = 0.5$~m/s, with time step $\Delta T = 0.02$~s and total duration $N = 100$. Moreover, $\kappa = 10 + 10j$, $\xi = 10$, and noise powers $\sigma_z^2 = \sigma_c^2 = -80$~dBm. Measurement noise coefficients are set to $\alpha_{\tau} = 1 \times 10^{-9}$ and $\alpha_{\mu} = 2 \times 10^3$.

\textit{2) Algorithm Parameters:} The input/output layer sizes of all networks match the state/action space dimensions, with a hidden layer of 128 neurons. We set the batch size to 512, discount factor $G = 0.99$, and policy-clipping factor $\epsilon = 0.2$~\cite{11091493}. Learning rates for actor and critic are $5\times10^{-5}$ and $5\times10^{-4}$, respectively. For SNNs, the LIF parameters are $\tilde{T} = 6$, $\eta = 3$, $\mathcal{U}_{\text{th}} = 1$, $\mathcal{U}^r = 0$, and $\lambda = 0.5$~\cite{10636728}.

\textit{3) Baselines:} We compare the proposed Spiking Actor-Critic (Spiking-AC) method with several state-of-the-art baselines, including Proximal Policy Optimization (PPO)~\cite{schulman2017proximal}, Deep Deterministic Policy Gradient (DDPG)~\cite{lillicrap2015continuous}, Deep Q-Network (DQN), and a Random policy.

\subsection{Simulation Results}
\subsubsection{Communication Performance}
\begin{figure}[t]
	\centering
	\begin{subfigure}[b]{0.5\linewidth}
		\centering
        \includegraphics[width=1.0\linewidth]{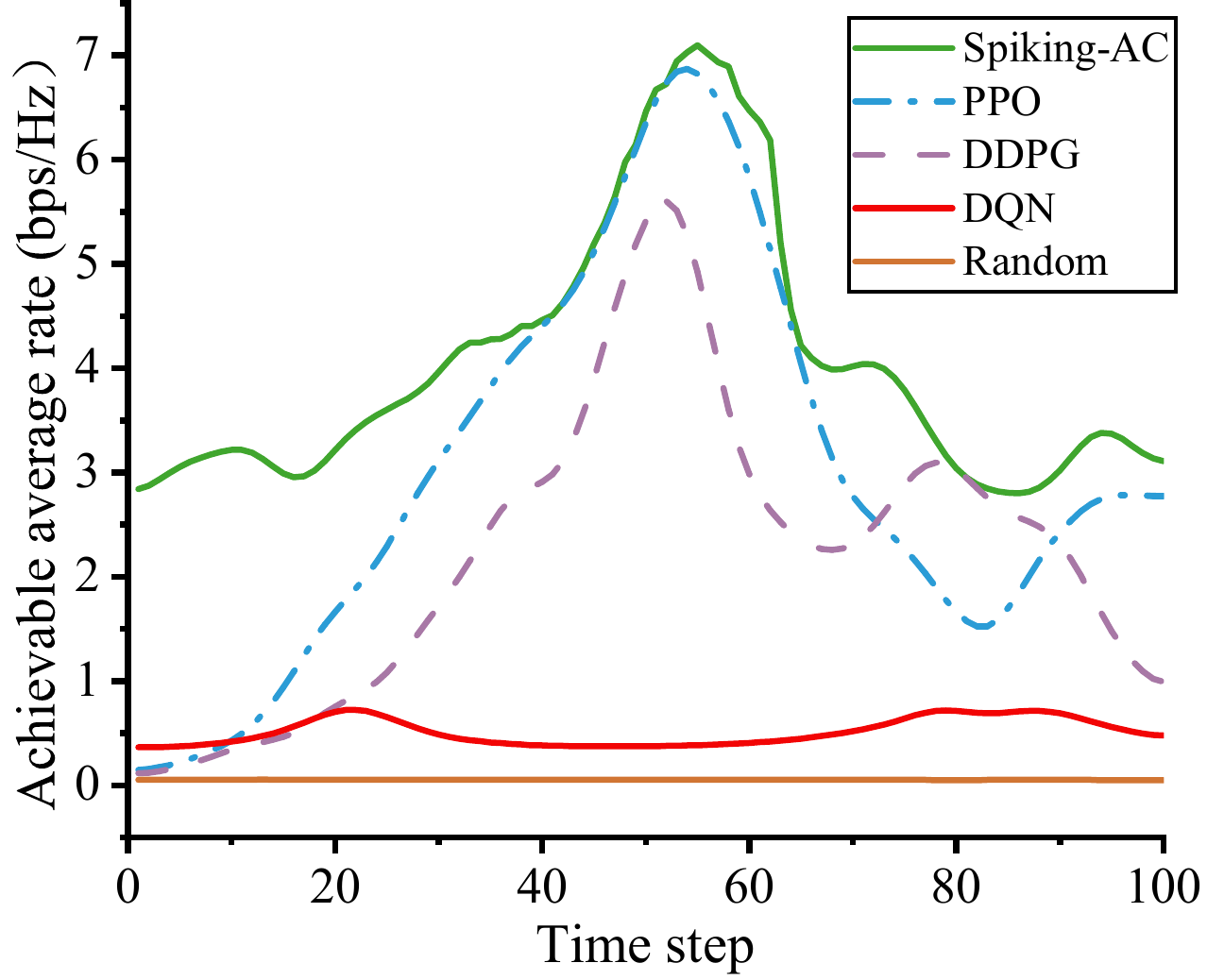}
		\caption{}
	\end{subfigure}%
	\begin{subfigure}[b]{0.5\linewidth}
		\centering
         \includegraphics[width=1.0\linewidth]{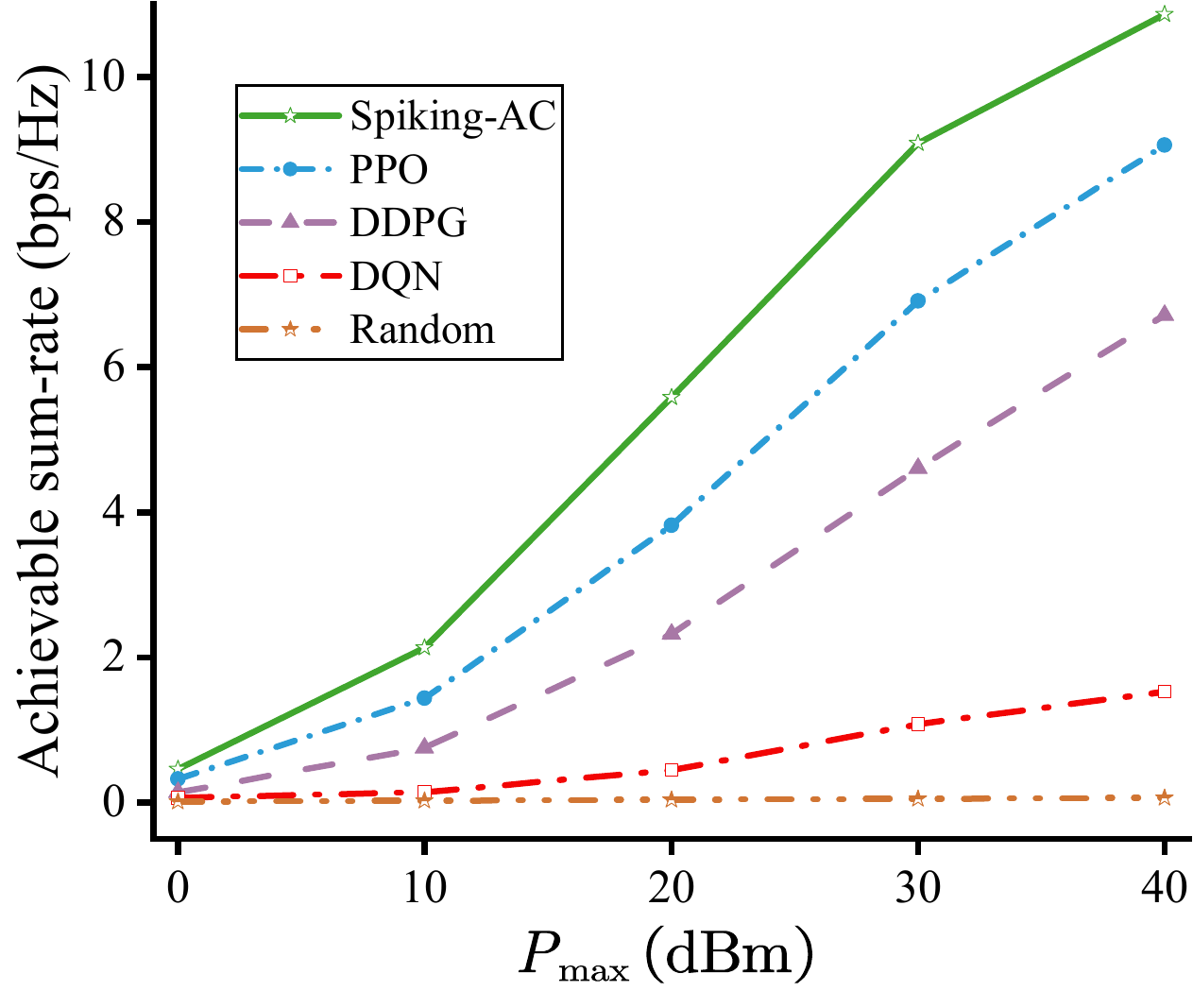}
		\caption{}
	\end{subfigure}
	\caption{(a) The achievable average sum-rate at $P_{\max}$ = 40 dBm and (b) the achievable sum-rate vs. maximum transmit power.} 
	\label{fig:communicationrate}
\end{figure}
We first evaluate the achievable communication rates during the vehicles' driving process. As shown in Fig.~\ref{fig:communicationrate}(a), Spiking-AC achieves the highest transmit rates in the first 40 time steps, demonstrating robust long-range communication performance. This is attributed to its efficient beamforming and power allocation, jointly optimized via SNN-based learning and the policy-clipping technique. By effectively mitigating path loss, Spiking-AC outperforms other methods that lack joint optimization capabilities. PPO, which shares the policy-clipping strategy, ranks second. DDPG, despite using an actor-critic framework, performs worse due to the absence of policy-clipping, leading to less stable training. As vehicles approach the RSU (time steps 40–60), all algorithms benefit from improved channel conditions. However, DQN shows a performance drop near the RSU, where interference management and resource coordination become critical. Its limited adaptability leads to degraded performance, highlighting the need for intelligent resource allocation in dense multi-vehicle scenarios.

We further assess performance under varying RSU power budgets by adjusting $P_{\max}$. As shown in Fig.~\ref{fig:communicationrate}(b), Spiking-AC consistently achieves the highest sum-rate across all power levels. At $P_{\max} = 40$~dBm, it attains 10.864~bps/Hz, outperforming PPO (9.058~bps/Hz) and DDPG (6.716~bps/Hz) by 19.96\% and 61.7\%, respectively. While performance gaps narrow under lower power budgets, Spiking-AC maintains superiority due to its energy-aware optimization strategy, ensuring robust communication even under strict constraints.
\subsubsection{Energy Consumption}
\begin{figure}[t]
	\centering
	\begin{subfigure}[a]{0.5\linewidth}
		\centering
        \includegraphics[width=1.0\linewidth]{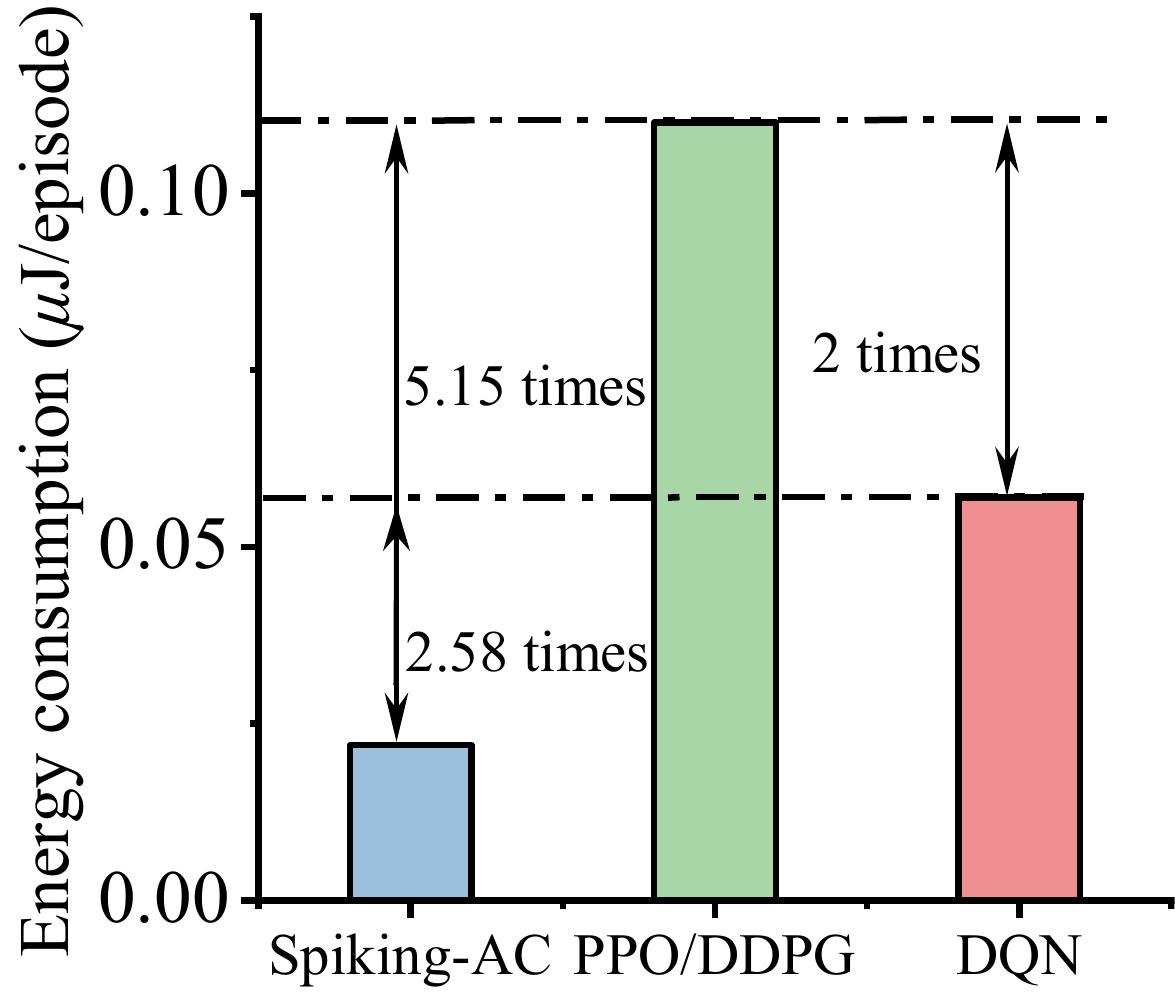}
		\caption{}
	\end{subfigure}%
	~
	\begin{subfigure}[a]{0.5\linewidth}
		\centering
        \includegraphics[width=1.0\linewidth]{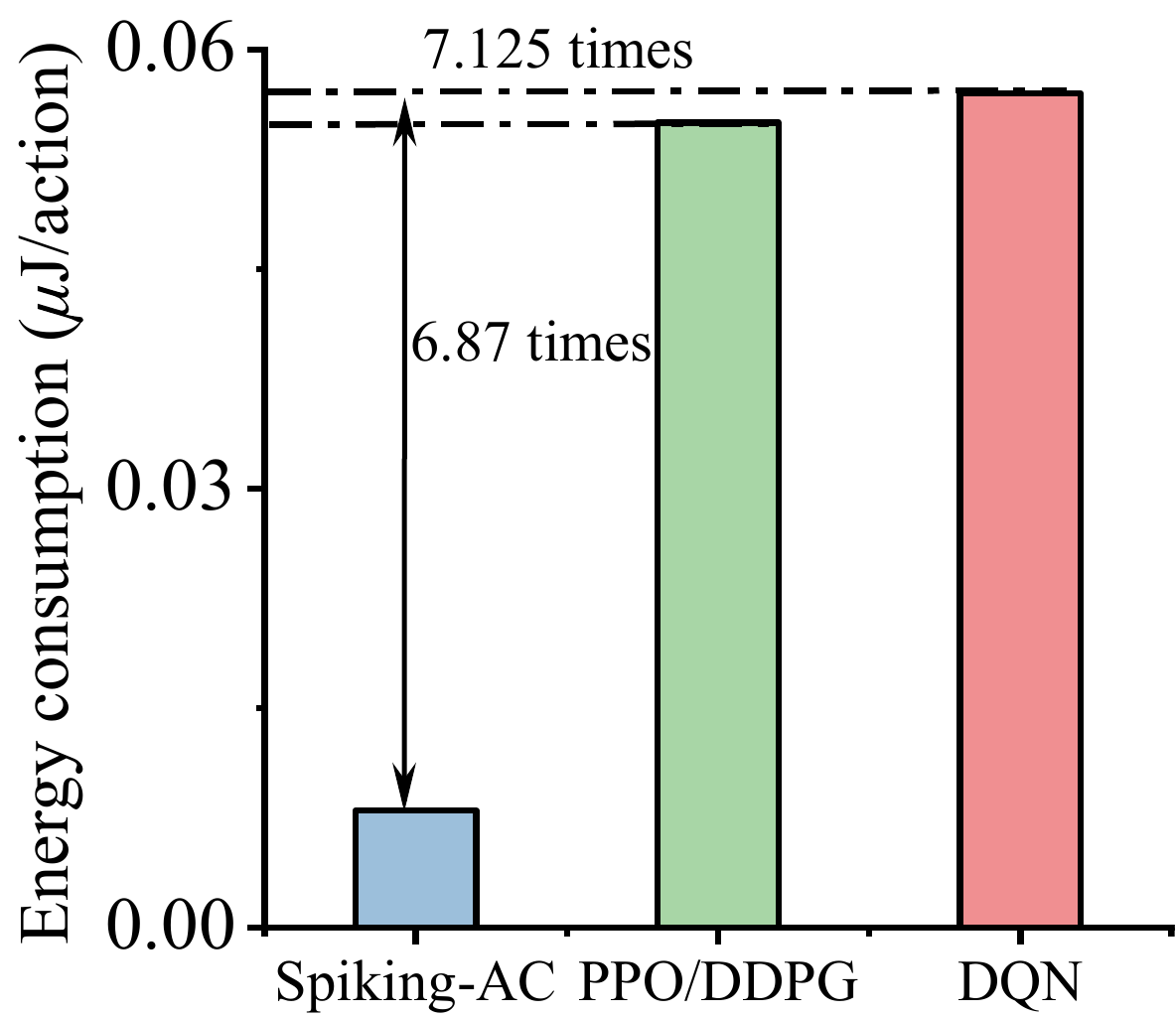}
		\caption{}
	\end{subfigure}
	\caption{(a) Comparison of energy consumption in training process and (b) comparison of energy consumption in inference process.}
	\label{fig:training-results}
    \vspace{-6pt}
\end{figure}
We then evaluate the energy consumption of different algorithms during model training and inference. To quantify this, we compute the total number of floating-point operations (FLOPs) and map them to energy consumption using hardware-specific coefficients for MAC and AC operations~\cite{panda2020toward,10636728}. As discussed in Section~\ref{algorithm}, Spiking-AC utilizes spike-based binary events and performs computation through AC operations, avoiding multipliers. In contrast, baseline methods rely entirely on energy-intensive MAC operations. For a three-layer network, the FLOPs for Spiking-AC and baselines are calculated as $FLOPS_{spiking}=\sum_{i=1}^2{N_i M_i \varsigma_i}$ and $FLOPS_{baseline}=\sum_{i=1}^3{N_i M_i}$, respectively, where $N_i$, $M_i$ are the layer dimensions, and $\varsigma_i$ is the spike firing rate~\cite{panda2020toward,10636728}.
The corresponding energy consumptions are~\cite{panda2020toward,10636728}:
$E_{spiking} = E_{AC} \cdot FLOPS_{spiking} \cdot \tilde{T} + E_{MAC} \cdot N_3 M_3$, 
$E_{baseline} = E_{MAC} \cdot FLOPS_{baseline}$, where $E_{AC} = 0.1$~pJ and $E_{MAC} = 3.2$~pJ, respectively.

As shown in Fig.~\ref{fig:training-results}(a) and (b), Spiking-AC reduces energy consumption by 5.15 times during training compared to PPO/DDPG, and by 2.58 times compared to DQN. During inference, the energy savings are even more significant due to the event-driven nature of SNNs. These results confirm that Spiking-AC enables efficient training and inference with over 50\% energy reduction compared to conventional learning-based methods, making it particularly suitable for energy-constrained V2X environments.
\addtolength{\topmargin}{0.02in}
It is worth to note that, the proposed Spiking-AC framework not only delivers significant communication performance gains through joint optimization of beamforming and power allocation, but also achieves remarkable energy efficiency by replacing energy-intensive MAC operations with lightweight spiking-based processing. This dual advantage highlights its potential for practical deployment in future green and intelligent vehicular networks.

\vspace{-5pt}
\section{Conclusion}\label{conclusion}
\vspace{-5pt}
In this work, we have proposed an energy-efficient learning-based beamforming algorithm for ISAC-enabled V2X networks. To this end, we have first modeled the dynamic and uncertain V2X environment as an MDP, allowing the RSU to adaptively perform beamforming and power allocation based on real-time sensing data. We have then employed a DRL algorithm with an actor-critic structure and policy clipping to learn robust control policies without relying on explicit CSI or prior noise statistics.
To further reduce energy consumption, we have integrated SNNs into the DRL algorithm, leveraging their event-driven nature for efficient training and inference. Simulation results have demonstrated that the proposed Spiking-AC algorithm outperforms baseline methods in communication rate and energy efficiency, offering a promising solution for sustainable V2X connectivity and green communication systems.

\vspace{-6pt}
\bibliographystyle{IEEEtran}
\bibliography{bibRef}
\vspace{-14pt}
\end{document}